\documentclass[usenatbib]{mn2e}
\usepackage{epsfig,journals}

\newcommand{\lesssim}{\lower.5ex\hbox{$\; \buildrel < \over \sim \;$}}
\newcommand{\gtrsim}{\lower.5ex\hbox{$\; \buildrel > \over \sim \;$}}
\newcommand{ \Ein }{ E_{\rm in} }
\newcommand{ \Rstar }{ R_\star }
\newcommand{ \RstarTZ }{ R_{\star \rm TZ} }
\newcommand{ \Mej }{ M_{\rm ej} }
\newcommand{ \tse }{ t_{\rm se} }
\newcommand{ \Ese }{ E_{\rm se} }
\newcommand{ \Tse }{ T_{\rm se} }
\newcommand{ \Msun }{ M_{\sun} }
\newcommand{ \Rsun }{ R_{\sun} }
\newcommand{ \lobster }{{\em LOBSTER}\ }
\newcommand{ \Cto }{ C_{2-1} }
\newcommand{ \Ctt }{ C_{3-2} }
\newcommand{ \NH }{ N_{\rm H} }
\newcommand{ \rhorho }{ \rho_1/\rho_\star }
\newcommand{ \deltat }{ \Delta t_{\nu, \gamma} }
\newcommand{ \tdwell }{ t_{\rm dwell} }

\begin{document}

\title[SN Properties from Shock Breakout]{Supernova Properties from
Shock Breakout X-rays}

\author[Calzavara \& Matzner]{Andrew J. Calzavara$^a$
\thanks{E-mail: andrew.calzavara@utoronto.ca; matzner@astro.utoronto.ca} 
and Christopher D. Matzner$^a$  \footnotemark[1] \\ $^a$Department of 
Astronomy and Astrophysics \\ University of Toronto, 60 St. George St., 
Toronto, Ontario, Canada, M5S 3H8}
\date{printed \today} 
\maketitle

\begin{abstract}
We investigate the potential of the upcoming \lobster space observatory
(due circa 2009) to detect soft X-ray flashes from shock breakout in
supernovae, primarily from Type II events.  \lobster should discover many
SN breakout flashes, although the number is sensitive to the uncertain
distribution of extragalactic gas columns.  X-ray data will constrain the
radii of their progenitor stars far more tightly than can be accomplished
with optical observations of the SN light curve.  We anticipate the
appearance of blue supergiant explosions (SN 1987A analogs), which will
uncover a population of these underluminous events.  We consider also how
the mass, explosion energy, and absorbing column can be constrained from
X-ray observables alone and with the assistance of optically-determined
distances.  These conclusions are drawn using known scaling relations to
extrapolate, from previous numerical calculations, the \lobster response
to explosions with a broad range of parameters.  We comment on a small
population of flashes with $0.2 < z < 0.8$ that should exist as transient
background events in {\em XMM}, {\em Chandra}, and {\em ROSAT}
integrations.  
\end{abstract}

\begin{keywords}
supernovae: general -- X-rays: bursts -- stars: fundamental parameters 
-- shock waves -- instrumentation: detectors.
\end{keywords}

\section{Introduction and Motivation}\label{Intro}
A core collapse supernova produces no electromagnetic
radiation until its envelope is completely consumed by the
explosion fireball.  This phase ends, however, with a brilliant flash
of X-ray or extreme ultraviolet photons heralding the arrival of
the fireball shock at the stellar surface.  The ``breakout''
flash is delayed in time, and vastly reduced in energy, relative to
the neutrino and gravity-wave transients produced by core
collapse.  However, it conveys useful information about the explosion
and the star that gave rise to it. 

We are motivated to explore the detection and interpretation of such
flashes by several points.  First, breakout flashes provide strong
constraints on some properties of their presupernova stars that are
poorly determined from the optical light curve.  Second, whereas
previous searches for these flashes have failed, the planned \lobster
experiment should routinely discover them.  Third, this experiment
will report (within minutes) the location of almost any core-collapse
supernova (some flashes are extincted) within its field of view to a
distance of many megaparsecs.  Data will be downlinked to Earth every 90 
minutes and analysed via an automated process.  This allows rapid 
optical follow-up and, in the unlikely case of a very nearby explosion, 
correlation with neutrino and gravity-wave signals.  Early warning and 
the precise timing of explosion are valuable for the interpretation of 
the optical light curve: for instance, in calibration of the Expanding
Photospheres Method for distance determination.  Fourth,
flash-selected surveys of Type II supernovae are less biased by
competition with light from galactic nuclei than are optical surveys.  
Finally, there exists the possibility that \lobster or a subsequent 
instrument could detect the signature of asymmetry in the supernova 
explosion through its effects on the breakout flash.

In this paper we consider primarily what can be determined about the 
stellar progenitor from the X-ray flash alone, from the flash and an 
optical distance, or from the delay separating the flash from the core 
collapse.  Our investigation is based on previous numerical calculations 
of the breakout flash spectrum, primarily by \cite{kc78}, \cite{ens92},
\cite{bli98}, and \cite{bli00}, and on the analytical approximations
and scalings derived by \cite{mat99}.  Our consideration of
observational points relies on technical details kindly provided by
Nigel Bannister and the \lobster science team.

\subsection{Limitations of Optical Light Curves for Constraining SN Properties}
\label{RelnToOpt}

Supernova properties provide a datum at the end of stellar
evolution.  Collectively, such data from many stars reveal how the 
intrinsic properties (e.g., initial mass, rotation, metallicity) and 
extrinsic properties (e.g., binary membership and separation) of massive 
stars affect their evolution.  It is currently uncertain, for instance, 
how many supernovae come from relatively compact blue supergiant
progenitors and therefore produce underluminous light curves like
that of SN 1987A \citep{1988slmc.proc..112S, 1988slmc.proc..106F, 
1988Natur.333..305Y, 1989ApJ...347L..29V, 1996ApJ...464..404S}.  This 
arises both from a bias against detecting small progenitors, and a 
bias to overestimate the radius, as discussed below.  X-ray 
observations have the potential to mitigate the former bias (Table 
\ref{t:ratesExt}) and eliminate the latter (\S \ref{constraints}). 

Using the scaling relations of \cite{ln85}, \cite{ham03} estimated
progenitor properties from the plateau properties of the 13
best-measured Type IIP SNe (the only ones for which sufficient
observations exist; prior to his work, only SN 1969C had been analyzed
in this fashion).  Based on measurement error, he assigned $1\sigma$
uncertainties of roughly 0.26 dex (a factor of 1.8) to the derived
radii (and similar uncertainties to mass and explosion energy).
Several additional sources of uncertainty degrade the determination of
progenitor radius derived from optical data. 

For one, there is an uncertainty in the underlying distribution of
ejecta density and opacity.  The scaling relations of \cite{ln85} are
derived from a suite of numerical simulations involving an explosion
in a simplified version of a red supergiant (RSG) model, in which the
progenitor density scales with radius as $\rho_0\propto
(1-r/\Rstar)^{1/8}$, where $\Rstar$ is the size of the star.
\cite{pop93}, in comparison, developed analytical formulae for
plateau-type light curves, assuming a constant density distribution in
the {\em ejecta}.  Although \citeauthor{ln85}'s results are nominally
more precise, being derived from a hydrodynamical calculation of a
model star, neither set of formulae account in detail for the
presupernova structure of the progenitor -- which depends, for
instance, on the mass ratio between the star's mantle and its hydrogen
envelope.  (Simple formulae for the ejecta distributions from
supergiants have been presented by \cite{mat99}, but have not yet been
used to predict light curves).  For a typical plateau light curve
\citep[absolute V magnitude peaking at -17.5$^{\rm mag}$ at 70 days;
early photospheric velocity 7000 km/s;][]{pop93}, the
\citeauthor{ln85} and \citeauthor{pop93} determinations of $\Rstar$
differ by 0.28 dex or a factor of 1.9, which is comparable to the
observational uncertainties.  This serves as an estimate (although
probably an overestimate) for the errors due to an unknown ejecta
distribution.

Additionally, there is  a bias toward large values  of $\Rstar$ due to
the  contamination of  the plateau  peak by  the radioactive  decay of
$^{56}$Co produced in the  explosion.  The formulae of \cite{ln85} and
\cite{pop93} assume the light curve to be powered by heat deposited in
the ejecta  by the explosion shock  itself.  At the  time that photons
can diffuse out of the ejecta, this heat supply has been adiabatically
degraded  during  expansion;   the  fraction  remaining  is  inversely
proportional to  the initial stellar radius.   In contrast, $^{56}$Co,
with a half-life of 77.3 days, releases its energy during the plateau.
If the initial radius is  small enough, cobalt decay will compete with
or  even dominate shock-deposited  heat.  To  detect this  effect, one
must monitor  the late-time decay of  the light curve  to estimate the
cobalt contribution.   Nevertheless, the derived  progenitor radii are
easily  corrupted, because  there is  no  simple way  to subtract  the
radioactive contribution  from the light curve; nor  is this possible,
if the  $^{56}$Co luminosity exceeds  that from shock  heating.  Using
the \cite{pop93} formulae, we find that radioactivity dominates when
\begin{equation} \label{CoLum}
\frac{M_{^{56}\rm Co}}{0.12\Msun} > 
\frac{E_{51}^{5/6} R_{500}^{2/3}}{\kappa_{0.34}^{1/3} M_{\rm ej,10}}
\exp\left(\frac{0.56 \kappa_{0.34}^{1/6} M_{\rm ej,10}^{1/2} R_{500}^{1/6}}
  {E_{51}^{1/6}} \right), 
\end{equation}
where $R \equiv 500 R_{500} \Rsun$ (an alternate, $R\equiv 50
R_{50}\Rsun$, will be used later for blue supergiants); the ejecta
mass is $\Mej \equiv 10 M_{\rm ej,10} \Msun$; the ejecta opacity is
$\kappa \equiv 0.34 \kappa_{0.34}$ cm$^2$ g$^{-1}$; and the explosion
energy is $\Ein \equiv 10^{51} E_{51}$ erg.  

For the fiducial red supergiant
($R_{500}=E_{51}=M_{10}=\kappa_{0.34}=1$), equation (\ref{CoLum})
indicates that $0.21\Msun$ or more of $^{56}$Co would be required to
dominate the plateau luminosity.  (This is a large but not impossible
amount: see \citeauthor{ham03} [2003]'s Table 4).  In contrast,
reducing the initial radius to $50\,\Rsun$, only $0.038\,\Msun$ of 
$^{56}$Co is needed to dominate the luminosity.  Clearly, one
cannot detect the existence of blue supergiant (BSG) progenitors from 
light curve plateaus if there is even a minute amount of radioactive cobalt.
This can also be seen by solving equation (\ref{CoLum}) for the value
of $R$ below which $^{56}$Co dominates: 
\begin{equation}\label{R-CoLum} 
\Rstar < 180 \frac{M_{10}^{0.35} \kappa_{0.34}^{0.34}}{E_{51}^{1.0}}
\left(\frac{M_{^{56}\rm Co}}{0.1\,\Msun}\right)^{1.34} \Rsun. 
\end{equation}
This local, power-law solution to the implicit equation
(\ref{CoLum}) is reasonably accurate for typical doses of
$^{56}$Co.  The scaling $M_{^{56}\rm Co}\propto \Rstar^{0.75}$ indicates
that $^{56}$Co contamination is an issue even for red supergiants. 

These difficulties prohibit the use of a catalogue of plateau-phase SN
light curves to derive the population of red versus blue supergiant
progenitors.  Further, such a catalogue would necessarily be biased by
the fact that larger stars and those with more $^{56}$Co have
intrinsically brighter optical displays.

There are other means for placing constraints on $\Rstar$ optically. 
One possibility is to locate the progenitor star in previous optical
observations and to estimate $\Rstar$ from the observed luminosity and
spectral type (see, for example, \citealt{van03}).  Of course, this
method hinges on whether the progenitor identity can be established
from archival data.

Alternatively, sufficiently early observations may catch the SN light
curve in the phase of self-similar spherical  diffusion, during which the
light curve is powered by shock-deposited heat escaping from the
highest-velocity ejecta \citep{1992ApJ...394..599C}. This theory
constrains the combination $\Rstar \Ein^{0.91} \Mej^{-0.40}$ rather
than $\Rstar$ alone.  For instance, an application to the early
observations of the Type Ib/c SN 1999ex \citep{2002AJ....124.2100S}
yields an estimate $\Rstar \simeq 7.0 E_{51}^{0.91} M_{\rm
ej,\,10}^{-0.40} \Rsun$.  To yield $\Rstar$, this method requires a
knowledge of the time of explosion, observations of the SN in its
first days, and estimates of the distance, explosion energy, and
ejected mass.

As we shall show, $\Rstar$ is the parameter most tightly constrained
by the breakout flash.  X-ray observations are thus independent of and
complementary to optical follow-up, and provide the early
warning needed for \cite{1992ApJ...394..599C}'s method. Indeed, X-ray
and optical constraints can be merged to provide a more complete
picture of the discovered SNe.

\subsection{Previous work}\label{prevwork}

Shock breakout flashes were predicted by \cite{1968CaJPh..46..476C} as 
a source for (the then undetected) $\gamma$-ray bursts.  

\cite{kc78} carried out radiation hydrodynamical calculations for the
explosions of red giant stars, predicting breakout flashes detectable
by the soft X-ray telescope {\em HEAO-1}.  Unfortunately, the field of
view of {\em HEAO-1} was insufficient to detect any breakout flashes
in the limited duration of the experiment \citep{kc79}.

The explosion of SN 1987A, and the realization that archival images of
its progenitor indicated a blue rather than a red supergiant star,
stimulated a reanalysis of supernova breakout flashes by \cite{ens92}
and, most recently, by \cite{bli98} and \cite{bli00}.  These studies
represent an increase in sophistication toward the full numerical
treatment of this complicated, radiation-hydrodynamic problem.  They
are consistent with the observed ionization state of circumstellar
gas around the SN 1987A remnant \citep{ens92}.

The physics of shock breakout and the accompanied flash have also been
treated more generally with analytical approximations for many
years. The hydrodynamical problem -- that of a shock wave accelerating
down a density ramp -- was solved in a self-similar idealization by
\cite{GF56}, \cite{S60} and \cite{1966ApJ...143...48G} and used by
\cite{1966ApJ...143..626C} in their study of supernova hydrodynamics.
After \cite{1968CaJPh..46..476C}'s initial suggestion, the physics of
the breakout flash was reexamined analytically by \cite{IN89} and
\cite{mat99}; see \S \ref{physics}.

We wish to tie together the numerical and analytical work on breakout
flashes to predict the detectability by \lobster of SNe of a wide
variety of characteristics.  This requires a method of scaling the
results of a numerical simulation to a different progenitor mass,
radius, or explosion energy (or, less importantly, envelope
structure).  This is made possible by the work of \citeauthor{mat99}
(1999; hereafter MM~99), who developed a general analytical
approximation that accurately predicts the speed of the explosion
shock front.  Evaluated at the stellar surface, and combined with
\cite{IN89}'s theory, this gives an estimate of the onset, duration,
luminosity, and spectrum in the flash.  Although numerical studies
predict the spectrum, for instance, more accurately, MM~99's formulae
provide the scaling laws required to generalize them.

\subsection{This Work} \label{thiswork}

In this paper we examine the ability of sensitive, wide-field X-ray
detectors (in particular, the forthcoming \lobster instrument) to
constrain the properties of supernovae through observation of the
shock breakout flash.  We begin, in \S \ref{physics}, with a brief
overview of the physical processes which generate the breakout flash.
Section \ref{constraints} describes, in detail, how the intrinsic
properties of a supernova and its progenitor star can be gleaned from
the characteristics of the X-ray flash.  We attempt to account for the 
distribution of Galactic and extragalactic interstellar X-ray absorption 
by employing Cappellaro et al. (1997)'s estimate of the distribution of 
optical extinctions toward supernovae; however this distribution is quite 
uncertain.  A brief discussion of the shock travel time and its usefulness 
in constraining parameters is included (\S \ref{neutrino}); unfortunately, 
\lobster is very unlikely to observe a supernova near enough for gravity 
waves and neutrinos to be detected.  We include a short discussion on the 
use of instruments other than \lobster for detecting breakout flashes 
(\S \ref{other}); we conclude that there may be unnoticed breakout flash 
detections in the archived images of the {\em XMM, Chandra}, and 
{\em ROSAT} X-ray telescopes. In \S \ref{discussion}, we summarize how 
shock breakout observation can be used as a powerful tool for 
understanding the end stages of stellar evolution and underline some 
caveats.  In the Appendix, we discuss, in greater detail, the significance 
of the outer density distribution of the progenitor and show how it can be 
characterized by other stellar parameters.  We model stellar luminosity as 
a function of stellar mass for presupernova stars and use this relation to 
estimate the outer density coefficients for blue supergiants.

Our analysis improves upon previous work by dealing more generally
with breakout flash properties and by modelling more specifically the
observation of bursts.  Previous work has concentrated on the
prediction of the breakout flash from a single star; the method of
scaling used in this work allows for us to describe flashes from a
broad range of progenitors (albeit in less detail).  We concentrate on
the observation of these events by \lobster and the reconstruction of
the supernova properties from the data.

\section{Physics of breakout flashes}\label{physics}

We briefly review here some of the physics relevant to breakout
flashes here; however, we refer the reader to \cite{IN89} and MM~99,
and references therein for a more thorough discussion.

MM~99's theory for the speed of the explosion shock front (their
Eq. [19]) reads 
\begin{equation} \label{vs} 
v_s(r) = 0.794 \left[\frac{\Ein}{m(r)}\right]^{1/2} 
  \left[\frac{m(r)}{r^3 \rho_0(r)}\right]^{0.19} 
\end{equation} 
where $\Ein$ is the explosion energy, $r$ labels position within the
progenitor, and $\rho_0(r)$ and $m(r)$ are the progenitor's density
and {\em ejected} mass interior to $r$.  This formula accounts in a
simple manner for two effects: deceleration as the shock sweeps up new
material (the first bracket), and acceleration in regions of sharply
declining density (the second bracket).  The prefactor is matched to a
self-similar solution; \cite{2001ApJ...551..946T} have given a
slightly more accurate form in which this coefficient depends on the
shape of the density distribution.  Likewise, the exponent 0.19 on the
second term can be adjusted (very slightly) depending on the slope of
the density profile near the stellar surface.  Nevertheless, equation
(\ref{vs}) is accurate within $10\%$, often $3\%$, as it stands.

The breakout flash itself occurs because supernova shocks are regions
of radiation diffusion, and when they get too close to the stellar
surface, photons diffuse out into space.  Such shocks are driven
predominantly by radiation rather than gas pressure, and this is
increasingly true as they accelerate through the diffuse subsurface
layers.  The shock thickness ($l_s$) is therefore determined by the
requirement that photons must diffuse upstream across the shock (in a
time $\sim \tau_s l_s/c$, where $\tau_s$ is the optical depth across
$l_s$) as fast as fluid moves downstream across it (in a time
$l_s/v_s$).  Equating these, 
\begin{equation}\label{taus}
\tau_s = \frac{c}{f\,v_s}.  
\end{equation} 
We have included a factor $f\sim 1-3$ which is uncertain
at this level of approximation because the shock is a smooth
structure; MM~99 assume $f=1$, arguing its effect is unimportant. 
We retain $f$ to show its effect, but note that it always appears
below as a coefficient modifying the X-ray opacity $\kappa$. 

The quantity $\tau_s$ decreases as the shock accelerates, but the
optical depth to the surface ($\tau$) decreases more rapidly.  Once
$\tau<\tau_s$, there is insufficient material to contain the shock,
and the photons that constitute it will escape and produce the flash.

To specify the flash's properties, one requires a formula for the
subsurface density profile; MM~99 motivate 
\begin{equation} \label{rho} 
\rho(r) = \rho_1 \left(\frac{\Rstar-r}{r}\right)^n, 
\end{equation} 
where $n$ is related to the polytropic index $\gamma_p \equiv d\ln
p/d\ln \rho$ via $\gamma_p = 1 + 1/n$. Hence $n\simeq 3/2$ in an
efficiently convecting atmosphere (i.e., in red supergiants) and
$n\simeq 3$ in a radiative atmosphere of uniform opacity (blue
supergiants).  Evaluating equation (\ref{rho}) at $r = \Rstar/2$
results in $\rho_1 = \rho(\Rstar/2)$; thus, $\rho_1$ represents the 
half-radius density.  The outer density distribution of a star can
be described by the ratio $\rhorho$, where $\rho_\star$ is the 
characteristic density $\Mej/\Rstar^3$ (see \S \ref{params} for more 
detail).  

Finally, one requires the relevant opacity $\kappa$ in the expression
$\tau = \int \rho \kappa \ dr$.  Since the post-shock temperature is
$\sim 10^5$ -- $10^6$ K, MM~99 assumed $\kappa$ would be dominated by
electron scattering --- hence, $\kappa\simeq 0.34$ cm$^2$ g$^{-1}$ for
solar metallicity.  We have investigated this assumption for the
relevant densities and temperatures using the OPAL tables
\citep{1996ApJ...464..943I}, and find it perfectly valid. 

The results of setting $\tau_s = \tau$ are given by MM~99 in their \S
5.3.3; we will not repeat them here.  Their formulae (36), (37), and
(38) give the characteristic temperature $\Tse$ (hence, photon
energy $\sim 2.7 k_B \Tse$), flash energy $\Ese$, and radiation
diffusion time $\tse$, respectively (the latter sets a lower limit
for the observed flash duration).  We do, however, wish to note
several important features of these formulae (Table \ref{t:MM99eqs}).

\begin{table*}
\centering
\begin{minipage}{190mm}
\caption{Scaling equations from MM~99. {These should be read as exponents
in formulae, e.g., $\tse = 10^{0.74} [f\kappa/(0.34$ cm$^2$
g$^{-1})]^{-0.58}$... cgs units.} \label{t:MM99eqs} } 
\begin{tabular}{@{}lrrrrrrr@{}}
\hline
  & \multicolumn{3}{c}{RSG} & & \multicolumn{3}{c}{BSG} \\
\cline{2-4} \cline{6-8} \\
Parameter & $\Tse$ & $\Ese$ & $\tse$ & & $\Tse$ & $\Ese$ & $\tse$ \\
\hline 
$10$ & $6.28$ & $46.49$ & $0.74$ & & $6.12$ & $46.88$ & $1.60$ \\
$f\kappa/0.34$ cm$^2$ g$^{-1}$ & $-0.10$ & $-0.87$ & $-0.58$ & &
$-0.14$ & $-0.84$ & $-0.45$ \\
$\rhorho$ & $0.070$ & $-0.086$ & $-0.28$ & & $0.046$ & $-0.054$ &$-0.18$ \\
$\Ein/10^{51}$ erg & $0.20$ & $0.56$ & $-0.79$ & & $0.18$ & $0.58$ &$-0.72$ \\
$\Mej/10 \Msun$ & $-0.052$ & $-0.44$ & $0.21$ & & $-0.068$ & $-0.42$ &$0.27$ \\
$\Rstar/50 \Rsun$ & $-0.54$ & $1.74$ & $2.16$ & & $-0.48$ & $1.68$ & $1.90$ \\
\hline
\end{tabular}
\end{minipage}
\end{table*}

First, these quantities depend most strongly on the progenitor radius
$\Rstar$, then, decreasingly, on $\Ein$, the total mass of ejecta ($\Mej$),
$\kappa$, and finally, quite weakly on the outer density coefficient
$\rho_1/\rho_\star$.  For this reason, we will ignore variations of the
structural parameter $\rho_1/\rho_\star$ in the current paper, and
instead concentrate our attention on $\Rstar$, the parameter to which
flashes are most sensitive.  We caution, however, that the effect of
$\rho_1/\rho_\star$ should be considered in cases where the outer
density profile can be determined.   An example would be a radiative
envelope of known luminosity; as discussed by
\cite{2001ApJ...551..946T}, $\rho_1/\rho_\star$ becomes very small if
the star's luminosity approaches its Eddington limit.  In the Appendix
we present alternative breakout scaling relations that incorporate 
analytical approximations for $\rhorho$. 

Second, there is a correlation between the photon temperature $\Tse$
and diffusion time $\tse$ induced by the physics of shock breakout.
Since photon pressure dominates, $\Tse$ must satisfy $a \Tse^4/3 =
(6/7) \rho v_s^2$ at the point of breakout.  And, since the diffusion 
time of the shock equals its crossing time, $\tse = l_s/v_s$.  These
expressions, combined with $\tau = \int \kappa \rho\ dr = l_s \rho
\kappa/(n+1)$ and $\tau = c/(f v_s)$, give
\begin{equation}\label{tseTse}
\Tse^4 \tse = \frac{18(n+1) c}{7 f a \kappa}. 
\end{equation}
As we have noted, $\kappa\simeq 0.34$ cm$^2$ g$^{-1}$ for envelopes of
roughly solar composition.  Insofar as $\tse$ and $\Tse$ can be measured,
equation (\ref{tseTse}) provides a means to determine observationally
whether the duration of an observed flash is dominated by diffusion
across the shock front.  

Alternatively, the flash duration may be dominated by the finite
difference in light travel time between the centre and the limb of the
stellar disk \citep{ens92}.  We estimate this as
\begin{equation}\label{tc}
t_c \simeq \frac{\Rstar}{c},
\end{equation}
although one should keep in mind that the emission pattern at
each point on the stellar surface (i.e., limb darkening) will make the
effective value of $t_c$ slightly smaller.  In general, the observed
flash is a convolution of the diffusion and propagation profiles;
hence the total duration is
\begin{equation}\label{tgam}
t \simeq (\tse^2 + t_c^2)^{1/2}, 
\end{equation}
although neither profile is actually Gaussian. 

Returning to equation (\ref{tseTse}), one sees that $\kappa$ will be
underestimated if one substitutes $t$ for $\tse$, provided $t_c > t$.
Likewise, the characteristic photon temperature can be increased
relative to $\Tse$ by selective absorption along the line of sight.
If one derives $\kappa\ll 0.34$ by substituting observed quantities in
Eq. (\ref{tseTse}), then either $t_c>\tse$, or extinction is
significant, or (quite likely) both.  Unfortunately, \lobster will not
view bursts long enough (\S \ref{lobster}) to apply this test in the
cases where it would be useful.  We concentrate below on other means
for constraining SN progenitors.

\section{Supernova Parameters from X-ray Constraints} \label{constraints}

There are five explosion parameters one would wish to know: the
intrinsic parameters $\Rstar$, $\Ein$, and $\Mej$; the distance $D$;
and the obscuring column $\NH$.  In addition, the structural parameter
$\rhorho$ plays a minor role.  As we shall see in the Appendix, 
$\rhorho$ is determined approximately by the stellar luminosity (in 
BSGs) or the mass fraction in the outer envelope (in RSGs).  In each 
case, however, $\rhorho$ is roughly constant at a characteristic value.

As described in \S \ref{physics}, the intrinsic parameters of the
explosion determine the characteristic properties of the breakout
flash: its colour temperature $\Tse$, its total energy $\Ese$, and its
duration $t$.  In theory, the full breakout spectrum provides a large
number of parameters.  However, we take the view that only these three
are independent.  If so, then it is not possible to reconstruct
$\Rstar$, $\Ein$, and $\Mej$ unless $\rhorho$ is held constant.
Variations of $\rhorho$ around its characteristic value leads to some
uncertainty in the other parameters.

The X-ray observables depend on $D$ and $\NH$ in addition to $t$,
$\Tse$, and $\Ese$.  Since there is negligible redshift at the
distances in question, $D$ affects only the X-ray fluence ($\propto
\Ese/D^{2}$ for given $\Tse$).  For this reason, $\Ese$ cannot be
known unless $D$ can be ascertained from optical followup
observations.

Given the $\sim 20\%$ uncertainty in {\em LOBSTER}'s determination of
photon energies \citep{ban03}, it provides roughly nineteen independent energy
channels (\S \ref{lobster}); with the burst duration, this makes
twenty observables per flash.  However, we use only four parameters in
our analysis: three channels (defined in \S \ref{method}), plus the
burst duration $t$.  

\begin{figure}
\centerline{\epsfig{figure=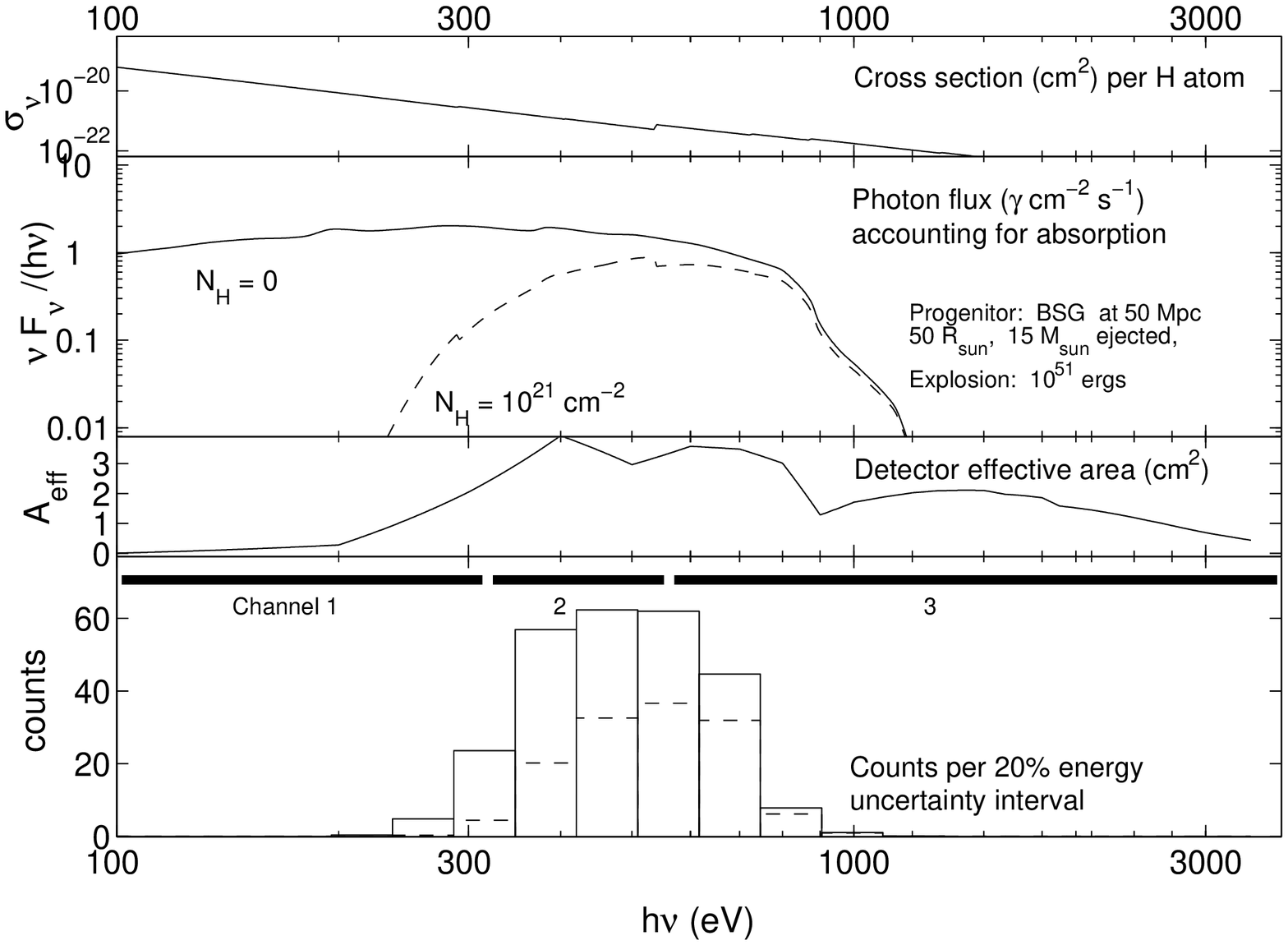,width=3.5in}} 
\caption[] {\footnotesize Construction of the \lobster instrument
  response to a blue supergiant SN at a distance of $2D_{\rm min} = $
  50 Mpc.  Eight events are expected at this distance
  (Eq. [\ref{Nobs}]).  Top panel: interstellar cross section per H
  atom \citep{wil00}.  Second panel: \cite{bli00}'s model for the
  breakout flash of SN 1987A, scaled with MM~99's scaling relations
  (solid) and extincted by a typical column (dashed).  Third panel:
  \lobster effective area.  Bottom panel: counts for the unextincted
  (solid) and extincted (dashed) flash, binned in 20\% bins (the
  energy uncertainty of the detector).  Also shown are the three energy
  channels we use to characterize colour. 
\label{f:schematicBlue} }
\end{figure}

\begin{figure}
\centerline{\epsfig{figure=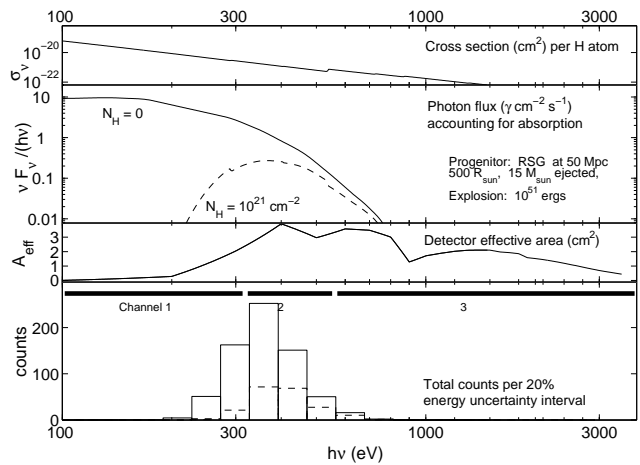,width=3.5in}} 
\caption[]{\footnotesize As in Figure \ref{f:schematicBlue}, but for a
  red supergiant model. The breakout flash was calculated by
  \cite{bli98} for SN 1993J; see \S \ref{method}. 
\label{f:schematicRed} }
\end{figure}

The motivation for this choice is twofold.  First, there are five
desired intrinsic parameters, two of which ($\Ese$ and $D$) cannot be
disentangled without optical followup; therefore only four X-ray
observables need be used.  Second, if supernovae occur homogeneously
within the detection volume, the number of bursts observed scales as
the limiting fluence according to $N \propto $(fluence)$^{-3/2}$.  The
median burst fluence is therefore only $60\%$ above the
detection threshold, and cannot be described in fine detail. 

After describing the \lobster instrument (\S \ref{lobster}), we
describe our method for simulating X-ray observables (\S \ref{method})
and our strategy for constraining supernova parameters with these
four observables (\S\S \ref{timing}--\ref{L-colour}). 

\subsection{Model of the \lobster instrument response} \label{lobster}

\begin{table*}
\centering
\begin{minipage}{140mm}
\caption{SN detection distance and rate versus model type.  The factor
  $f_{\rm obs}$ represents the unknown population of each subclass
  (Eq. [\ref{Nobs}]).  In each model, $\Mej = 15 \Msun$, $\Ein = 10^{51}$
  erg, and $\kappa = 0.34$ cm$^2$ g$^{-1}$, unless otherwise noted.
\label{t:rates}}
\begin{tabular}{@{}lcrrrl@{}}
\hline
Progenitor  & $\NH$ & $\Rstar$ & $D_{\rm max}$ & Detections & Note \\
Type  & ($10^{21}$ cm$^{-2}$) & ($\Rsun$) & (Mpc) & per year & \\ 
\hline
BSG              & 1 & 50  & 260 & $\sim 320 f_{\rm obs}$ & 
   SN 1987A analogue \\ 
BSG              & 5 & 50  & 86  & $\sim 12 f_{\rm obs}$ &
   SN 1987A analogue \\ 
RSG              & 1 & 500 & 420 & $\sim 1400 f_{\rm obs}$ & \\
RSG              & 5 & 500 & 66  & $\sim 6 f_{\rm obs}$ & \\
RSG              & 1 & 500 & 800 & $\sim 10 000 f_{\rm obs}$ &
   SN 1993J analogue; $\Mej = 2 \Msun$ \\ 
Wolf-Rayet core  & 1 & 5   & 58  & $\sim 1 f_{\rm obs}$ &
   BSG scaled to SN Type Ib/c analogue; \\
& & & & & $\Mej = 3 \Msun$, $\kappa = 0.2$ cm$^2$ g$^{-1}$, 
$\rhorho = 0.1$ \\
\hline
\end{tabular}
\end{minipage}
\end{table*}

\lobster is a proposed wide-field soft X-ray detector, currently
envisioned to be deployed on the International Space Station in 2009.
\lobster has a field of view of 162$\times$22.6 degrees and would
continually scan the sky as it orbits the Earth every 90 minutes.
Doing this, it would image \gtrsim 95\% of the sky once every 
orbit.  \lobster has a spatial resolution of
approximately 4 arcmin, and is sensitive to soft X-rays of energy 0.1
-- 3.5 keV from sources with fluxes of order $10^{-12}$ erg cm$^{-2}$
s$^{-1}$ \citep{pri96}.  Incident photon energies are recorded with an
accuracy of roughly $20\%$, implying $\sim 19$ effective spectral
channels.

\lobster focuses X-rays through grazing angle reflections by an array
of microchannels.  This arrangement produces a cruciform focal spot on
the detector, with a bright central focus spot at the intersection of
two dimmer, orthogonal arms; there is also a diffuse background of
unfocused photons.  The instrument's effective area is a complicated
function of incident photon energy, being limited at low energies by
detector window absorption and at higher energies by decreased
reflectivity.  Shown in the third panel of Figure
\ref{f:schematicBlue} is the effective area for the central
focus, from data provided by Nigel Bannister (2003, private
communication).  For a full discussion, see \citet{pri96}.

The instrument's field of view should allow for successful
detection of numerous shock breakouts.  One can define the 
instantaneous volume of view
\[ V_{\rm view} = \frac{1}{3} \Omega D_{\rm max}^3 \]
where $D_{\rm max}$ is the maximum distance at which a certain flash
can be detected.  The field of view, $\Omega$, is a narrow swath of
the celestial sphere: $\Omega=2 (162^\circ) \sin(22.6^\circ/2) = 1.11$
sr.  

The instantaneous volume of view is appropriate for events that are
shorter than the dwell time $\tdwell$ of the experiment.  The
dwell time for \lobster has a minimum value of roughly (90
min)(22.6$^\circ$ /360$^\circ$) = 340 s, with longer times occurring
near the poles of the orbit, to a maximum of $\tdwell = 1800$ s
\citep{ban03}.  We will adopt the typical value of 400 s for
$\tdwell$.  Longer events can be found even if their beginnings are
not observed, leading to a larger effective solid angle $\Omega_{\rm
eff} \simeq (1+t/t_{\rm dwell})\Omega$, until the event is viewed in
more than one scan of the sky (although this is not expected for
breakout flashes).  This increase in sensitivity for long
events comes at a cost: the burst duration $t$ is no longer directly
observable, and can only be constrained.  

Note that \cite{kc78} assumed that breakout flashes would outlast the
30 s dwell time of {\em HEAO 1}, and calculated their detection rate
accordingly.  {\em LOBSTER}'s dwell time, in contrast, is an order of
magnitude longer; only RSGs (with $\Rstar > 170 R_{\sun}$) have $t > 
\tdwell$ for this experiment.  

\cite{cap99} quote a rate of $0.71 h^2$ Type II supernovae per century 
per $10^{10}$ solar luminosities in the B band.  There are  $2\times 
10^8 h L_\odot {\rm Mpc}^{-3}$ in the B band \citep{fuk98}; from 
\citet{spe03} we adopt $h=0.71$ (the Hubble constant in units $100\, 
\rm km\, s^{-1}\, Mpc^{-1}$).

If a fraction $f_{\rm obs}$ of Type II supernovae belong to a subclass
of interest, and if this subclass is observable to distance $D_{\rm
max}$, then the number of flashes of this type expected in the total
integration time of the instrument's life, $t_{\rm obs}$, is
\begin{equation} \label{Nobs}
N_{\rm obs} = f_{\rm obs} \left[\frac{D_{\rm max}}{D_{\rm min}(t_{\rm obs})}
\right]^3. 
\end{equation}
The quantity $D_{\rm min}(t_{\rm obs})$, a function of the duration of the
experiment, is the minimum distance at which one would expect,
statistically, to observe one SN II of any progenitor type:
\begin{equation}
D_{\rm min}(t_{\rm obs}) = 26 \left[ \frac{3\, \rm
  yr}{t_{\rm obs}} \left(\frac{\tdwell}{t+t_{\rm
  dwell}}\right)\right]^{1/3}
 {\rm Mpc}. 
\end{equation}
Although this is comparable to the distance to the Virgo cluster, the
universe is reasonably homogeneous on scales $D>D_{\rm min}$. (The
local overdensity may reduce the actual minimum distance by $\sim
25\%$ relative to $D_{\rm min}$; however we neglect this effect in its
definition.)

For the \lobster mission, $t_{\rm obs} \approx 3$ years; Table
\ref{t:rates} shows $D_{\rm max}$ and $N_{\rm obs}$ for various
progenitor types, including the canonical SNe II (a RSG progenitor with 
$\Mej = 15 \Msun$, $\Rstar = 500 \Rsun$, $\Ein = 10^{51}$ erg, and a
BSG progenitor with $\Mej = 15 \Msun$, $\Rstar = 50 \Rsun$, 
$\Ein = 10^{51}$ erg).  Although the number of events observed depends
strongly on parameters, \lobster should observe between $\sim
10^2$ and $10^3$ SNe II and of order one SN Ib/c over a three year
lifespan.  In \S \ref{abs} we refine these estimates by averaging
over the distribution of absorbing columns $\NH$; see Figure
\ref{f:DmaxvsNh} and Table \ref{t:ratesExt}.

Note that in Table \ref{t:rates} we have included Type Ib/c
supernovae, which occur five times less frequently than Type IIs
\citep{cap99} and hence have $D_{\rm max} = 44$ Mpc.  Their
observability is rather strongly dependent on the rather poorly known
initial radius, ejected mass, and explosion energy: for these events,
$N_{\rm obs} \propto (\Ese/\Tse)^{3/2} \propto \Ein^{0.4} \Mej^{-0.4}
\Rstar^{2}$ (quite approximately, because $A_{\rm eff}(h\nu)$ is not
flat and extinction is not completely negligible).  In the table, we
denote by $f_{\rm obs}$ an unknown fraction with the properties given
in Table \ref{t:rates}.  

\subsection{Prediction of Breakout Flash Properties} \label{method}

We describe in this section the prediction of X-ray observables from
the properties of a model explosion.  We begin by calculating the
intrinsic parameters $\Ese$, $\Tse$, and $\tse$ using the equations
presented by MM~99 as shown in Table \ref{t:MM99eqs}.  We choose from
the literature a numerical calculation of the breakout spectrum as
close as possible to the desired explosion, then shift the spectrum to
make its total energy and mean photon energy conform to these model
predictions.

Unfortunately, there are very few published calculations of breakout
flashes.  For red supergiants there is the calculation by \cite{kc78},
and the more recent model for SN 1993J \citep{bli98}.
\citeauthor{kc78}'s model includes a hard tail of higher-energy
photons that is absent in \citeauthor{bli98}'s model, presumably due
to the more sophisticated radiation transfer employed in the latter.
Because \cite{bli98} employ multi-group transfer calculations, we
adopt this calculation as the fiducial RSG flash.
However, there is insufficient information about the progenitor in the
literature to calculate $\Tse$ for it using the MM~99 equations.
When using it, therefore, we enforce that the mean photon energy
agrees with the value $2.7 k\Tse$ appropriate for a blackbody of
that colour temperature.

For blue supergiants there are two potential sources of breakout spectra: 
\cite{ens92} and \cite{bli00}, both of which were calculated for SN
1987A.  We use the latter, again because the multigroup radiation
transfer method employed therein is the more sophisticated.  
In this case the MM~99 formulae could be applied to the progenitor
model, and predict, within 20\%, the mean photon energy.  We
nevertheless enforce a mean energy of $2.7 k\Tse$ when
constructing model BSG flashes, to be consistent with the RSG case.

For all Type II supernovae we fixed $f\kappa$ at 0.34 cm$^2$
g$^{-1}$ (\S \ref{physics}).

\begin{table}
\centering
\caption{Parameters used in scaling breakout spectra. \label{t:param}}
\begin{tabular}{@{}lccc@{}}
\hline
Progenitor & $\Mej (\Msun)$ & $\Rstar (\Rsun)$ & $\rhorho$ \\
type & & & \\
\hline
BSG & $1$ -- $35$ & $20$ -- $100$ & $0.2$ \\
RSG & $1$ -- $35$ & $50$ -- $1000$ & $0.5$ \\
\hline
\end{tabular}
\end{table}

Our three spectral channels were defined as follows: 0.10 -- 0.33 keV
(channel 1), 0.33 -- 0.54 keV (channel 2), and 0.54 -- 3.5 keV
(channel 3).  The chosen energy range matches the response of the
\lobster instrument, 0.1 -- 3.5 keV (\S \ref{lobster},
Figs. \ref{f:schematicBlue} and \ref{f:schematicRed}).  The lower
bound of channel 2 was chosen such that all flashes had sufficient
channel 1 photons at $D = D_{\min}$ after suffering the expected
amount of interstellar extinction ($\NH \sim 10^{21}$ cm$^{-2}$); the
upper bound was set so that all RSG flashes had at least some channel
3 photons prior to extinction.

If $C_{1,2,3}$ are the counts in the three channels, we may define
colour parameters $\Cto = \log_{10}(C_2/C_1)$ and $\Ctt =
\log_{10}(C_3/C_2)$. In the absence of an independent distance
determination (see the introduction to \S \ref{constraints}), these
two colour parameters and the flash duration $t$ are the only
constraints on the explosion itself.  Unlike $t$, $\Cto$ and $\Ctt$
are affected by absorption.

Figures \ref{f:schematicBlue} and \ref{f:schematicRed} illustrate the
prediction of the \lobster instrument response to typical blue and red
supergiant explosions, respectively. 

\subsubsection{Interstellar X-Ray Absorption}\label{abs}

\begin{table*}
\centering
\begin{minipage}{140mm}
\caption{SN detection rate averaged over a distribution of
 extinctions. SN models from Table \ref{t:rates}, the Galactic column
 density distribution \citep{1998ApJ...500..525S} and an estimate of
 the extragalactic column distribution (Fig. 4 of \citealt{cap97}) are
 used to find the typical distance (Eq. \ref{<D_max>_abs}) observable
 number, and column density (averaged over observable flashes,
 Eq. \ref{<logNH>_abs}) toward each type; see Fig. \ref{f:DmaxvsNh}.
 These are sensitive to the small-$\NH$ tail of the extragalactic
 column distribution, which we reiterate is quite uncertain; see
 Figure \ref{f:DmaxvsNh}. 
\label{t:ratesExt}}
\begin{tabular}{@{}lcccc@{}}
\hline
Type  &  $\Rstar$ & $\left<D_{\rm max}^3\right>^{1/3}$ &
Detections & $\left<\log_{10} \NH ({\rm cm}^{-2})\right>$ \\
 & ($\Rsun$) & (Mpc) & per year &  \\ 
\hline
BSG (87A analogue) &  50  & 96  & $ 17 f_{\rm obs}$ & 21.40
\\ 
RSG                &  500 & 113 & $27  f_{\rm obs}$ & 21.32
\\
RSG (93J analogue)   &  500 & 223 & $210 f_{\rm obs}$ & 21.33
\\ 
W-R core (Ib/c)    &  5   & 37  & $0.2 f_{\rm obs}$ & 21.68
  \\
\hline
\end{tabular}
\end{minipage}
\end{table*}

In the X-ray, bound-free absorption by the interstellar medium (ISM)
is greatest for the softest photons.  This leads to a ``bluening'' of
the X-ray spectrum as opposed to reddening in the optical.  To account
for the effects of absorption, we apply the X-ray opacities of
\citet{wil00} and consider a range of hydrogen column densities $\NH$.
Milky Way columns \citep{1998ApJ...500..525S} can roughly be described
with a log-normal distribution: $\log_{10}(\NH/{\rm cm}^{-2})\simeq
10^{20.72\pm0.62}$ (1$\sigma$ error bars).  

Host galaxy column densities toward Type II and Type Ib/c supernovae
are quite uncertain; however there is evidence they are often
significantly larger than typical Milky Way columns.  \cite{cap97}
estimate blue magnitudes of extinction toward SNe of both types by
searching for trends with galaxy inclination (after accounting for the
type of the host and adopting a fiducial value for the face-on
extinction).  Adopting the model plotted in their Figure 4
leads\footnote{We quote columns in H atoms cm$^{-2}$, but these are
derived from magnitudes of extinction in B, and applied to X-ray
absorption.  The end result should therefore be roughly independent of
host galaxy metallicity.}  to a range of $\NH$ from $1.5\times
10^{21}$ cm$^{-2}$ to $1.3\times 10^{22}$ cm$^{-2}$, in a distribution
strongly skewed toward higher values.   

These are comparable to the columns through Galactic molecular clouds,
which may reflect the interstellar context in which they explode.  On
the other hand, molecular gas is seen to disperse from OB associations
on timescales shorter than the typical presupernova lifetime
\citep{1991psfe.conf..125B}.  Moreover, \cite{1991PASAu...9...13V}
argues that chimneys around OB associations produce an observed
bias toward face-on galaxies.  In these two ways, energetic feedback
from OB stars may reduce $\NH$ in a fraction of the supernova
population. 

\begin{figure}
\centerline{\epsfig{figure=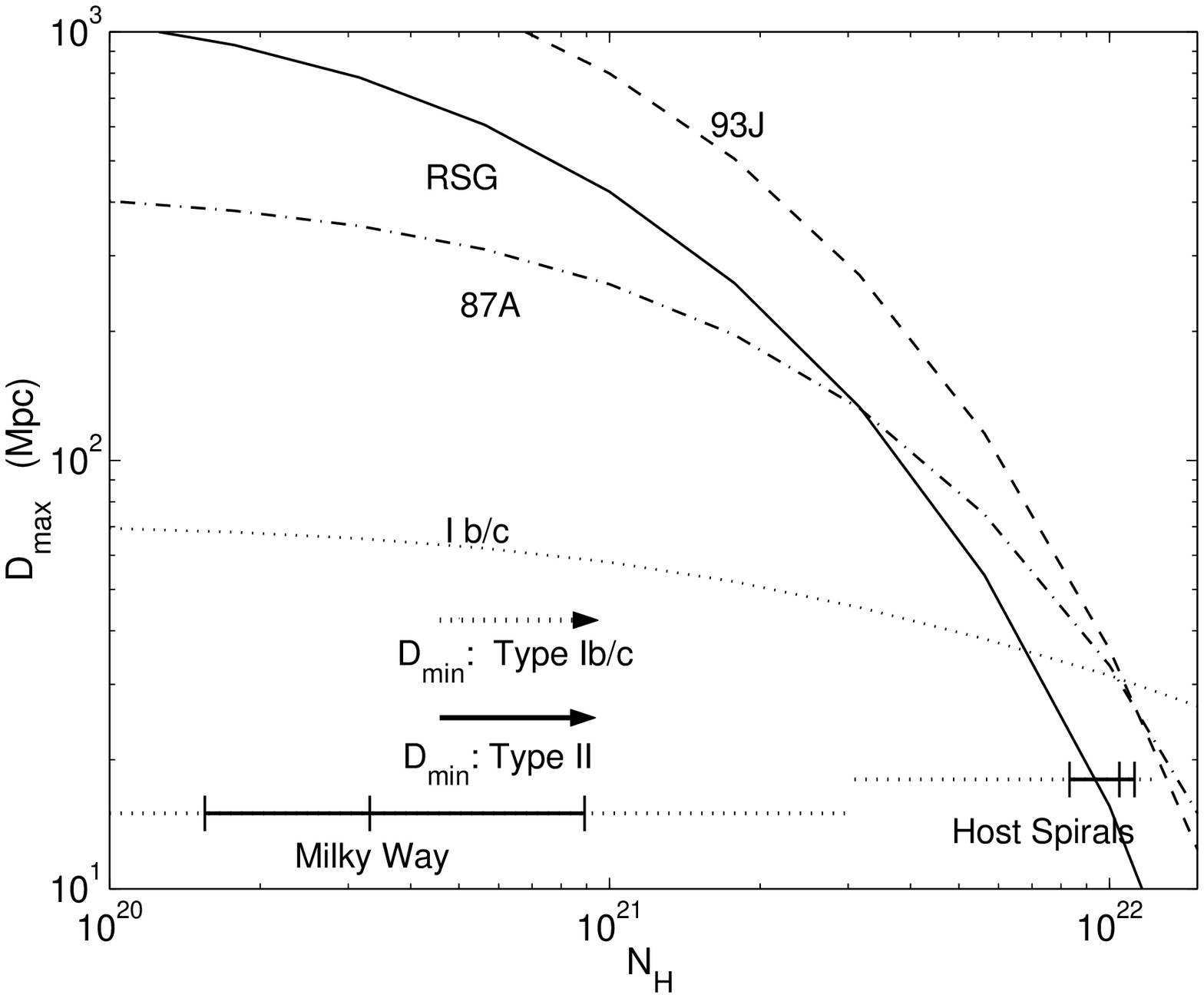,width=3.5in}} 
\caption[] {\footnotesize Maximum distance of observability, $D_{\rm
max}$, versus absorbing column $\NH$ for several progenitor models.
Models with harder spectra are less sensitive to extinction.  Also
plotted are the distances $D_{\rm min}$ within which one supernova is
expected in three years; one expects $\sim D_{\rm max}^3/D_{\rm
  min}^3$ events.  The distribution of columns due to the Milky
Way \citep{1998ApJ...500..525S} and inferred for host spiral galaxies
\cite{cap97} are shown (dotted lines: 10th to 90th percentiles; solid
dashes: 25th, 50th, and 75th percentiles).  Note that the host column
distribution is very uncertain. 
\label{f:DmaxvsNh} }
\end{figure}

The Galactic and extragalactic column distributions are shown in
Figure \ref{f:DmaxvsNh}.  Since $N_{\rm obs}\simeq D_{\rm
max}^3/D_{\rm min}^3$, it is quite clear that the statistics of the
\lobster breakout-flash catalogue will be sensitive to the low-$\NH$ end
of the host-galaxy column distribution.  This is demonstrated in Table
\ref{t:ratesExt}, which gives the effective value of $D_{\rm max}$
averaged over the extinction distribution: 
\begin{equation}\label{<D_max>_abs}
\langle D_{\rm max}^3 \rangle \equiv \int D_{\rm max}(N_H)^3
\frac{d{\cal P}(<\NH)}{d\NH} d\NH,  
\end{equation}
where ${\cal P}(<\NH)$ is the probability that the column density is less
than $\NH$.  The expected number is then $N_{\rm obs} = f_{\rm obs}
\langle D_{\rm max}^3\rangle /D_{\rm min}^3$ as in Eq. (\ref{Nobs}). 

Table \ref{t:ratesExt} also gives the log-normal value of $\NH$,
averaged over the observed bursts:
\begin{equation}\label{<logNH>_abs}
\langle \log_{10}\NH \rangle \equiv \frac{\int D_{\rm max}(N_H)^3
\frac{d{\cal P}(<\NH)}{d\NH} \left(\log_{10}\NH\right) d\NH}
{\int D_{\rm max}(N_H)^3
\frac{d{\cal P}(<\NH)}{d\NH} d\NH}. 
\end{equation}
Although these numbers are subject to our uncertainty of ${\cal P}(<\NH)$, 
they give a sense of what may realistically be observed in a survey.
Both estimates strongly weight the low end of the $\NH$ distribution, 
causing the `typical' column to be only $\sim 2-2.5\times 10^{21}$ 
cm$^{-2}$.

The stellar wind of the progenitor star can have a column far in
excess of the interstellar values: for a mass loss rate $\sim 10^{-5}
M_{\sun}/$yr and a terminal velocity of order the escape velocity, it is
$\sim 2\times10^{23}(10 M_{\sun}/M_{\star})^{1/2} (50 R_{\sun}/\Rstar)^{1/2}
{\rm cm^{-2}}$.  However, this wind is fully ionized by a small
fraction of the breakout photons \citep[see][]{1988A&A...192..221L}
and cannot significantly affect the X-ray spectrum.

The closest supernova expected in the \lobster catalogue is located
$D_{\rm min} \approx 26$ Mpc away, as defined in equation (\ref{Nobs})
assuming three years of observation (RSGs with $t > t_{\rm dwell}$ 
should be detected at a shorter distance; \S \ref{lobster}).  If the 
total number of counts
detected from a supernova at $D_{\rm min}$ is less than 10 (a fiducial
number), we consider the supernova {\em unobservable}.   (Unobservable
flashes have $D_{\rm max}<D_{\rm min}$.)  Likewise, we
consider the colour parameters $\Cto$ and $\Ctt$ {\em uncharacterized}
if there are not enough counts ($< 5$ in a necessary channel) to
construct them given a SN at $D_{\rm min}$.  These definitions are
employed in Figures \ref{f:tvsC32}, \ref{f:tvsC21}, \ref{f:C-C}, and
\ref{f:L-C} to identify excessive values of $\NH$. 

When the effective area of the \lobster instrument is taken into account,
analysis reveals that all BSG and RSG progenitor models with $\Ein = 
10^{51}$ erg are unobservable when $\NH > 10^{22}$ cm$^{-2}$.  The colour 
parameter $\Cto$ is uncharacterized for all models when $\NH > 2 \times 
10^{21}$ cm$^{-2}$ and $\Ctt$ is uncharacterized when $\NH > 8 \times 
10^{21}$ cm$^{-2}$.

\subsection{Constraints From Timing Alone} \label{timing}

It is possible to place constraints on the progenitor radius simply 
by measuring the duration of the shock breakout burst and comparing 
the light travel time to the diffusion time for the given range of 
parameters in Table \ref{t:param}.  

One can define a zone of transition from light travel time dominated
flashes to those dominated by diffusion.  By setting $t_c = \tse$ 
and solving for $\Rstar$, one obtains an expression for the radius 
of a RSG progenitor which is in the transition zone ($\RstarTZ$):
\begin{eqnarray}\label{RstarTZ}
\RstarTZ &=& 
697 
\left(\frac{f \kappa}{0.34 {\rm cm}^2 {\rm g}^{-1}}\right)^{0.50}
 \left(\frac{\rho_1}{\rho_\star}\right)^{0.24} 
\nonumber\\ &\times&
 \left(\frac{\Ein}{10^{51} {\rm erg}}\right)^{0.68}
\left(\frac{\Mej}{10 \Msun}\right)^{-0.18}
{\Rsun}.
\end{eqnarray}
The duration of a flash from a transition zone supernova ($t_{\rm TZ}$) 
is simply $\RstarTZ / c$:
\begin{eqnarray}\label{tTZ}
t_{\rm TZ} &=&
27.0
\left(\frac{f \kappa}{0.34 {\rm cm}^2 {\rm g}^{-1}}\right)^{0.50}
 \left(\frac{\rho_1}{\rho_\star}\right)^{0.24} 
\nonumber\\ &\times&
 \left(\frac{\Ein}{10^{51} {\rm erg}}\right)^{0.68}
\left(\frac{\Mej}{10 \Msun}\right)^{-0.18}
{\rm min},  
\end{eqnarray}
slightly longer than the \lobster dwell time. 

Hence, for breakout bursts with duration $t < t_{\rm TZ}$, the
progenitor radius is $ct$, and for those with longer duration,
$\Rstar$ can be constrained via the $\tse$ definition of MM~99.  
Similar equations can be derived to define the transition zone for BSG
progenitors, though it should be noted that, for the parameters considered
(Table \ref{t:param}), the breakout flashes from BSGs are always light 
travel time dominated when $\Ein \gtrsim 10^{51}$ erg; hence, a 
transition zone exists only for BSG explosions with less than this
canonical energy.

As an example, consider a supernova with $\Ein = 10^{51}$ erg which produces 
a breakout flash with a well-measured duration $t$.  Using the parameters 
given in Table \ref{t:param} and equation \ref{tTZ}, one can establish 
minimum and maximum values for the light travel and diffusion times of RSG 
and BSG progenitors and draw conclusions from how $t$ falls into these ranges.  
If $t < 116$ s, the progenitor is a BSG and its radius is $ct$ (since no 
RSG progenitor has such a small light travel time).  For a duration of 
116 -- 230 s, the colour of the flash could be used to distinguish between 
RSG and BSG (see \S \ref{time-colour}), and the progenitor radius is $ct$ 
(since both RSGs and BSGs can have a $t_c$ of this magnitude).  
A flash $>230$ s long denotes a RSG (since no BSG is large enough to produce
a flash this long); for 230 s $< t <$ 1100 s the radius is $ct$ (since
the minimum value of $t_{\rm TZ}$ is 1100 s), and for $t > 2080$ s 
restrictions can be placed on $\Rstar$ using the equation for $\tse$.  
Flashes with duration 1100 -- 2080 s are RSG type and could be either 
$t_c$ or $\tse$ dominated (since the maximum value of $t_{\rm TZ}$ is 2080 s).
Note, however, that the duration of a flash is not well constrained by
\lobster if it exceeds the $\sim 400$ s dwell time of the experiment. 

These results are summarized in Table \ref{t:timing}.  

\begin{table}
\centering
\caption{Constraints on progenitor type and $\Rstar$ from timing
  alone. An explosion energy $\Ein = 10^{51}$ erg is assumed.  Note
  that 1100 -- 2080 s is the region spanned by the transition zone 
  (TZ); hence, a breakout flash with this duration could be either 
  $t_c$ or $\tse$ dominated -- more information is required.}
\label{t:timing}
\begin{tabular}{@{}cll@{}}
\hline
$t$ (s) & Progenitor & $\Rstar$ \\
  & type & \\
\hline
$0$ -- $116$ & BSG & $ct$ \\
$116$ -- $230$ & need colour info & $ct$ \\
$230$ -- $1100$ & RSG & $ct$ \\
$1100$ -- $2080$ & RSG & near TZ \\
$> 2080$ & RSG & can constrain using $\tse$ \\
\hline
\end{tabular}
\end{table}

\subsection{Constraints From Timing and Colour} \label{time-colour}

By considering the colour of a flash in addition to its duration, 
more information about the supernova can be deduced and certain
degeneracies may be broken.  For instance, a flash with duration 116 s
$< t <$ 230 s (from a $10^{51}$ erg supernova) is light travel time
dominated, but both large BSGs and small RSGs produce flashes of such
duration.  The colour of the flash may help to break this degeneracy
and allow for an identification of the progenitor type.

\begin{figure}
\centerline{\epsfig{figure=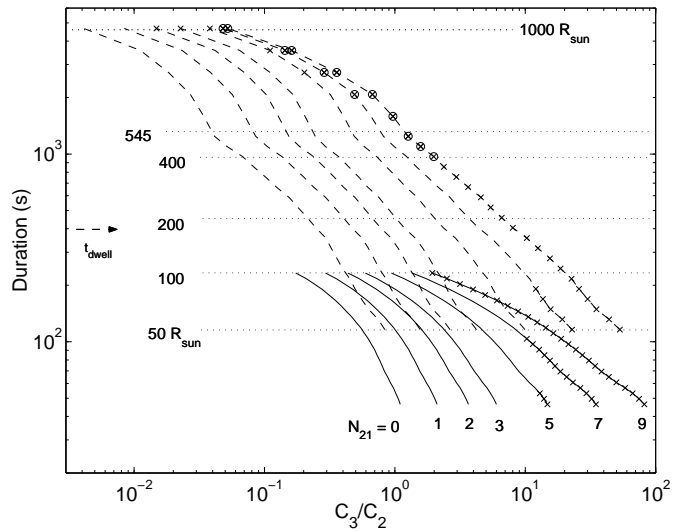,width=3.5in}} 
\caption[] {\footnotesize Duration and $\Ctt$ colour index for all
progenitor models; the solid line indicates BSGs and the dashed line
RSGs.  Each curve corresponds to a value of $\NH$, which increases to
the right in units of $10^{21}$ cm$^{-2}$ as indicated.  Dotted lines
mark constant radius.  Models marked with ``x'' are uncharacterized and 
those with ``$\circ$'' are unobservable (\S \ref{abs}).  
The discontinuity in slope in the RSG plot denotes the transition 
from light travel time to diffusion time in the burst duration.  Note 
that lines of constant radius for $t > t_{\rm TZ}$ are horizontal only 
because $\Mej$ and $\Ein$ are held constant in each model at 
$15 M_{\sun}$ and $10^{51}$ erg, respectively. 
\label{f:tvsC32} }
\end{figure}

\begin{figure}
\centerline{\epsfig{figure=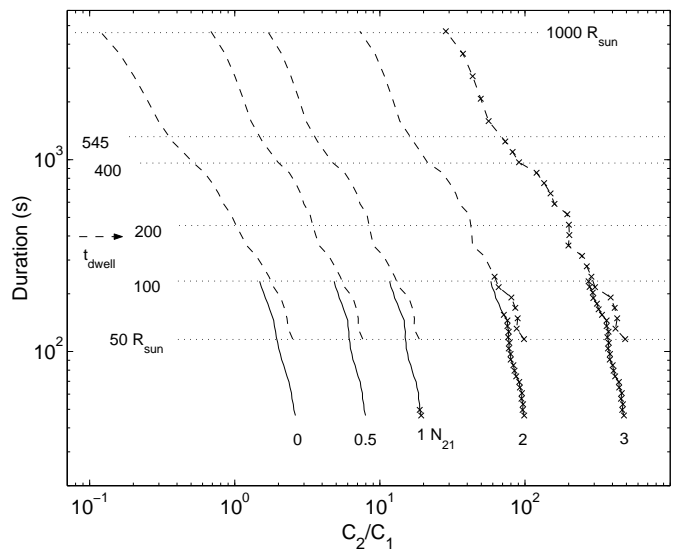,width=3.5in}} 
\caption[] {\footnotesize Duration and $\Cto$ colour index for all progenitor 
models; the solid line indicates BSGs and the dashed line RSGs.  Each curve
corresponds to a value of $\NH$, which increases to the right in units of
$10^{21}$ cm$^{-2}$ as indicated.  Dotted lines mark constant radius.  Points 
marked with ``x'' indicate models which are uncharacterized (\S \ref{abs}).  
The discontinuity in curvature of the RSG plot denotes the transition from 
$t_c$-dominated to $\tse$-dominated duration.  Note that lines of constant 
radius for $t > t_{\rm TZ}$ are horizontal only because $\Mej$ and $\Ein$
are held constant in each model at $15 M_{\sun}$ and $10^{51}$ erg, 
respectively. 
\label{f:tvsC21} }
\end{figure}

$\Tse$ varies inversely with $\Rstar$; hence, larger progenitors
produce redder flashes of lower radiation temperature with more
channel 1 photons than smaller progenitors do.  As can be seen in Figure
\ref{f:tvsC32}, a RSG produces a flash which is distinctly bluer in
colour than that of a BSG of the same radius (same flash
duration); however, RSG flashes in general are redder and brighter
since RSGs typically have much larger radii than BSGs.  The colour
difference between BSG and RSG flashes of the same radius is enhanced
by increasing the absorbing column density, as the redder BSG flashes
lose proportionally more channel 3 counts than their RSG counterparts
(this effect is evidenced by the divergence of the two types in $\Ctt$
space).  Figure \ref{f:tvsC21} shows that RSG and BSG flashes lose
proportionally the same amount of counts in channels 1 and 2 as
absorption increases (no divergence).  It seems unlikely that the
``type degeneracy'' mentioned in \S \ref{timing} can be broken without
an estimate for $\NH$ and an accurate measure of the flash colour.

As previously noted, increased absorption causes the values of $\Cto$ and 
$\Ctt$ to increase as lower energy photons are preferentially absorbed.
A greater $\Ein$ also leads to increased colour values due to its 
proportionality with $\Tse$, whereas $\Mej$ is inversely proportional to the 
radiation temperature (c.f. Table \ref{t:MM99eqs}).  The effect of $\Ein$ on 
colour is very small ($\sim 0.2$ power-law dependence) compared to that of 
$\NH$ (exponential dependence), and the effect of $\Mej$ is minute ($\sim 
-0.05$ power law dependence); $\Ein$ and $\Mej$ do have a significant effect
on $\tse$.  In the case of bursts with $t > \tdwell$, whose durations 
cannot be determined by \lobster, the SN properties are best constrained
using flash luminosity and colour (\S \ref{L-colour}).

Accurate observations of a flash's duration and colour can pinpoint
the location of the flash on a $t$ vs. $\Ctt$ or $t$ vs. $\Cto$ plot
(Fig. \ref{f:tvsC32} or \ref{f:tvsC21}) and hence constrain the value
of $\NH$ in addition to $R$.  If $\NH$ is known from other
observations, it can be used in tandem with \lobster colour
observations to constrain $\Mej$ and $\Ein$.

\begin{figure}
\centerline{\epsfig{figure=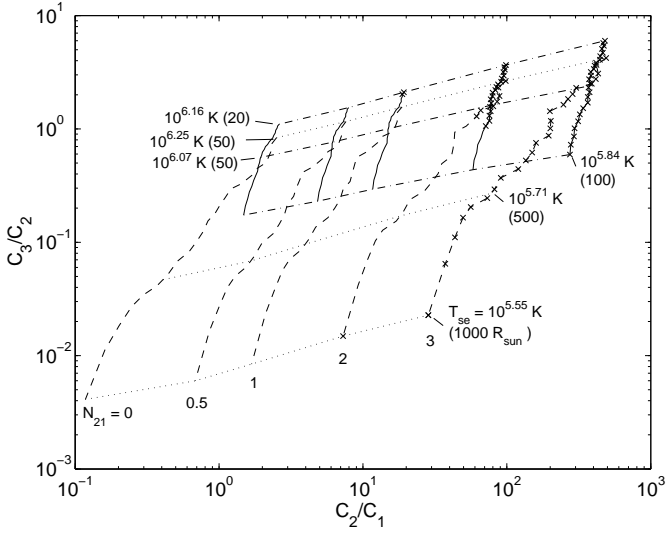,width=3.5in}} 
\caption[] {\footnotesize Colour-colour plot for all progenitor
models; the solid line indicates BSGs and the dashed line RSGs.  Each
solid or dashed curve corresponds to a value of $\NH$, which increases
to the right in units of $10^{21}$ cm$^{-2}$ as indicated.
Dash-dotted lines mark constant colour temperature $\Tse$ for BSG
models, dotted lines mark constant $\Tse$ for RSG models.  With each
$\Tse$ is provided (in parentheses) the value of $\Rstar$ at which
this value is attained in the case $\Ein=10^{51}$ erg and $\Mej=15
\Msun$.  Points marked with ``x'' indicate models which are
uncharacterized (\S \ref{abs}).
\label{f:C-C} }
\end{figure}

Potentially better constraints come from a colour-colour diagram.
Figure \ref{f:C-C} shows that the bluening effect of extinction can,
to some extent, be disentangled from the intrinsic flash colour to
provide an estimate of both $\NH$ and $\Tse$.  The latter can then be
taken as a constraint on $\Ein$, $\Mej$, and $\Rstar$, which is
significantly more illuminating if $\Rstar$ is constrained by other
means.  If, in addition, $\Ese/t$ is determined by comparing the
observed count rate to a value of $D$ (from optical followup;
Figs. \ref{f:L-C} and \ref{f:L-CBSG}; \S \ref{L-colour}), then all
explosion parameters are constrained. 

\subsection{Constraints from Luminosity and Colour}\label{L-colour}

Flashes that outlast the \lobster dwell time have durations that can
only be constrained rather than measured; this puts a lower limit of
roughly 170 $\Rsun$ on the progenitor.  Their durations may be either
diffusion or light travel-time dominated (\S \ref{physics}), but
equation (\ref{tseTse}) is unlikely to discriminate between these
possibilities with only a lower bound on $t_{\rm se}$.  

\begin{figure}
\centerline{\epsfig{figure=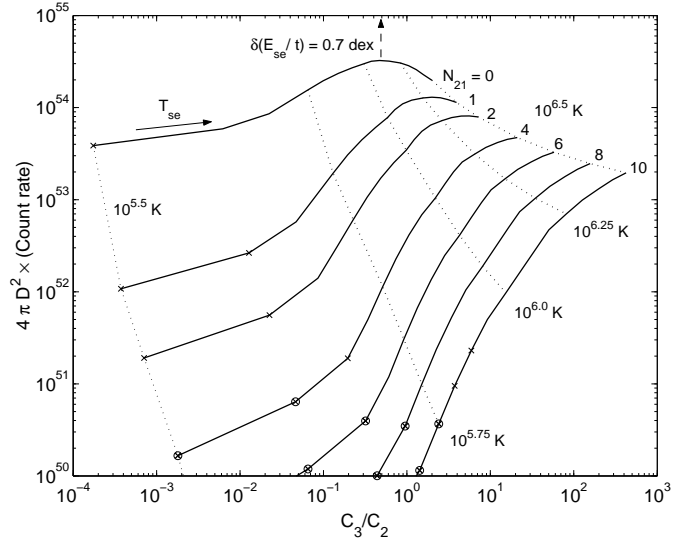,width=3.5in}} 
\caption[] {\footnotesize A theoretical X-ray Hertzsprung-Russell
  diagram of breakout flashes, in which luminosity (measured via the
  rate of observable photon production) and colour (here $\Ctt$) are
  compared for RSG models.  These quantities are observable even if 
  the burst is not
  viewed for its entire duration, and thus are useful for flashes with
  $t > \tdwell$.  Each curve corresponds to a value of $\NH$,
  which increases in units of $10^{21}$ cm$^{-2}$ as indicated.
  Dotted lines mark constant values of $\Tse$ as indicated; $\Tse$
  increases to the right along each curve.  All models are evaluated
  at constant $\Ese / t$ ($10^{45}$ erg/s); a change in
  $\Ese/t$ results in a vertical shift of the curves (the dashed
  arrows shows a 0.9 dex shift).  Models marked with ``x'' are
  uncharacterized and those with ``$\circ$'' are unobservable (\S
  \ref{abs}).  
\label{f:L-C} }
\end{figure}

The readily observable quantities for these long bursts are their
X-ray colours and brightnesses.  We will assume that optical follow-up
allows a determination of flash distances: in that case, one can
construct an X-ray Hertzsprung-Russell (H-R) diagram for them (Figure
\ref{f:L-C}).  In the H-R diagram, the rate of observable photon
production ($4 \pi D^2$ [Count rate]) is plotted versus a colour index.
Specifying $\Tse$, $\Ese / t$, and $\NH$ produces a point on the plot.
We hold $\Ese / t$ constant and vary $\Tse$ and $\NH$ to illustrate
trends; varying $\Ese / t$ simply results in a vertical shift of the
curves.

\begin{figure}
\centerline{\epsfig{figure=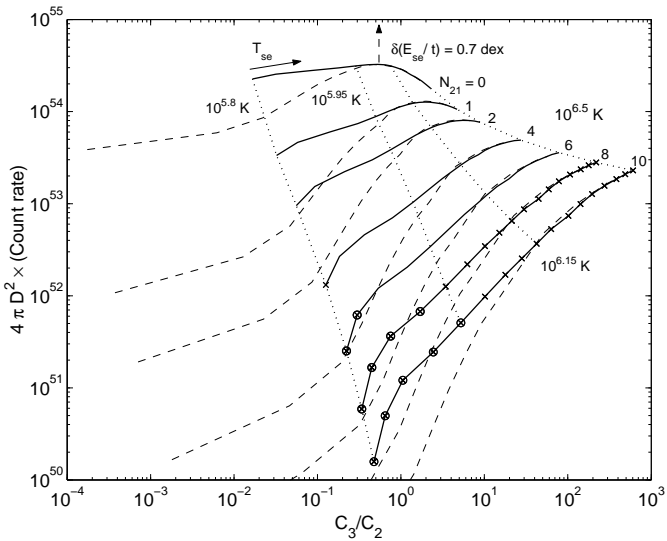,width=3.5in}} 
\caption[] {\footnotesize As in fig. \ref{f:L-C}, but for BSG models
  (thick solid lines).  RSG models appear as thin dashed  
  lines for comparison.  These quantities are observable even if 
  the burst is not viewed for its entire duration, although all BSG
  progenitors are expected to have duration $t < \tdwell$.  Each 
  curve corresponds to a value of $\NH$, which increases in units of 
  $10^{21}$ cm$^{-2}$ as indicated.  Dotted lines mark constant values 
  of $\Tse$ as indicated; $\Tse$ increases to the right along each curve.
  All models are evaluated at constant $\Ese / t$ ($10^{45}$ erg/s); a 
  change in $\Ese/t$ results in a vertical shift of the curves (the 
  dashed arrows shows a 0.9 dex shift).  Models marked with ``x'' are
  uncharacterized and those with ``$\circ$'' are unobservable (\S
  \ref{abs}).  
\label{f:L-CBSG} }
\end{figure}

By placing a point on Figure \ref{f:L-C} representing the observed
characteristics of a flash, it possible to fix $\Tse$.  
With other information, such as that provided by the colour-colour
diagram (Fig. \ref{f:C-C}), $\NH$ and $\Ese / t$ can be constrained.  
Through the scaling equations (\ref{t:MM99eqs}), $\Mej$, $\Ein$, and 
$\Rstar$ can, in turn, be constrained.  Although all BSG progenitors
are expected to produce flashes with $t < \tdwell$, we include
an H-R diagram for BSGs for the purposes of comparison (Fig. 
\ref{f:L-CBSG}).  For higher $\Tse$ models, BSG flashes are effectively
indistinguishable from RSG flashes of the same flash luminosity, $\Ese/t$.
The H-R diagram does not present a good means of constraining progenitor
type for flashes with $t < \tdwell$; the other methods described
in this section work best.  For long flashes which are not observed in
their entirety, the H-R diagram could prove invaluable.

\section{Shock travel time} \label{neutrino}

The travel time for the shockwave responsible for the breakout flash, or 
equivalently, the time lag between the emergence of neutrinos and gravity 
waves from the exploding star and the emergence of shock breakout, is well 
approximated by 

\begin{eqnarray} \label{eq:neutrino}
\deltat &=& \int_0^{R_\star} \frac{dr}{v_s} 
 \\ &=& 1.26  \frac{\Rstar\Mej^{1/2}}{\Ein^{1/2}} 
\int_0^1 \left[\frac{m(r)}{\Mej}\right]^{1/2}
\left[\frac{\rho_0(r) r^3}{m(r)}\right]^{0.19} \frac{dr}{\Rstar}. 
\nonumber 
\end{eqnarray}
Stellar mass and density profiles are needed to define $v_s$
(Eq. \ref{vs}); the authors used progenitor structures provided by
Woosley and Nomoto.  For a $16 M_{\sun}$, $48 R_{\sun}$ BSG with $\Mej
= 14 M_{\sun}$ and $\Ein = 10^{51}$ erg, $\deltat$ is 2 hours.  For an
$18 M_{\sun}$, $450 R_{\sun}$ RSG with $\Mej = 16.5 M_{\sun}$ and
$\Ein = 1.1 \times 10^{51}$ erg, the delay between neutrino/gravity
wave emission and shock breakout is 36 hours.  It should be noted that
this method to determine the time of shock emergence using equation
(\ref{eq:neutrino}) is considerably more accurate than the Sedov
solution for a constant density envelope, used by \citet{woo02} and
others.  The determination of $\deltat$ is not of consequence
observationally unless the supernova in question occurs within a few
kpc; in this case, it is feasible to detect the neutrino/gravity wave
emission with the appropriate detectors. Near-future gravity wave
detectors will only be able to detect SNe within our Galaxy
\citep{ott04}, whereas neutrino signals can be detected out to the
Magellanic clouds.  Neutrino and/or gravity wave detection would
provide: 1. early warning that a breakout flash (and supernova) will
occur, and 2. a measurement of $\deltat$, which could be used to
further constrain the supernova properties.  It should be noted,
however, that the rate for Galactic supernovae is very low ($\sim
0.01$ yr$^{-1}$), that only $\sim 9\%$ of these fall within the
\lobster field of view, and that most of these occur in the Galactic
disk, where $\NH > 10^{22}$ cm$^{-2}$ typically; hence, constraining
SNe properties using shock travel time could only be done under very
fortuitous circumstances.

\section{Shock breakout observation with other instruments} \label{other}

In examining the capabilities of an instrument to observe the shock
breakout flash, this paper has focused on the proposed \lobster
wide-field X-ray detector; however, breakout flashes are also
detectable by other current instruments -- though less frequently and
in less detail.  Preliminary work by the authors indicates that the
EPIC-PN instrument of the {\em XMM-Newton} X-ray space observatory, with 
a narrower field of view but much higher effective area than {\em LOBSTER}, 
is capable of detecting flashes at moderate redshifts.  The $\sim 0.2$ 
deg$^2$ field of view increases $D_{\rm min}$ (\S \ref{lobster}), or 
equivalently, $z_{\rm min}$, to $z_{\rm min} = 0.2$; but the high 
effective area allows detections of the canonical flashes to $z_{\rm max} 
\sim 0.9$ (c.f. $D_{\rm max}$, \S \ref{lobster}).  The {\em Chandra} X-ray
observatory's HRC-I has a similar field of view, and hence $z_{\rm min}$, as
{\em XMM Newton}, with a lower effective area and correspondingly lower
$z_{\rm max} \sim 0.2$.  The PSPC instrument on the {\em ROSAT} satellite, 
with a large field of view and intermediate effective area, has 
$z_{\rm min} = 0.1$, $z_{\rm max} \sim 0.6$.  The assumed columns are
$\NH = 2 \times 10^{21}$ cm$^{-2}$ in the host galaxy and $\NH = 0.5 \times 
10^{21}$ cm$^{-2}$ locally.  These results indicate that 
there may be a number of undiscovered breakout flashes in the backgrounds 
of archived {\em XMM Newton} and {\em ROSAT} images, of order 10 for each 
instrument.  In the HRC-I images, there may be of order 1 BSG breakout 
flash and likely none from RSGs; it should be noted that \citet{rac03} 
have come to a similar conclusion with regard to the ACIS instrument on 
{\em Chandra}.  

Because the breakout flash durations are much shorter than the typical 
integration times for these instruments, and because a typical flash 
detection would be dimmer than the X-ray sources observed, the probability
of a serendipitous discovery is almost zero.  This may explain why the 
archived breakout flashes have gone unnoticed, even if present in the data.
An in-depth search of the archived images would be required to locate the 
detected flashes. 

A sensitive optical survey with a sufficiently large field of view may be
able to detect the Rayleigh-Jeans tail of the breakout flash spectrum. 
Preliminary calculations indicate that, to an extinctionless limiting 
magnitude of $\sim 23$ mag in both the U and B bands (Johnson-Morgan 
system), there is one flash per square degree per $200 f_{\rm obs}^{-1}$ 
years for the canonical RSG, and one flash per square degree per $600 
f_{\rm obs}^{-1}$ years for the canonical BSG.  Extinction in the host
galaxy typically causes the limiting magnitude to increase to $\sim 24$ 
mag.

We have ruled out the possibility of detecting high redshift breakout
flashes with optical telescopes; the high luminosity distances involved
more than compensate for the cosmological K-correction.

\section{Discussion} \label{discussion}

We have combined an analytical theory for the dependence of shock
breakout parameters (MM~99) with numerical simulations of shock
breakout flashes \citep{kc78,bli98,ens92,bli00} to predict the
expected signal observed by the \lobster spaceborne soft X-ray camera.
Our emphasis has been on the reconstruction of
supernova parameters -- primarily radius, mass, explosion energy, and
obscuring column -- from the data (\S \ref{constraints}).

Supernova radius is the most tightly constrained of the three
quantities (\S \ref{timing}).  For all events which have a duration 
that can be measured individually, because it is shorter than the dwell 
time of the instrument, the duration is set by the light travel time of 
the star rather than the leakage of photons from the surface layers of 
the explosion.  These events come from stars with $R\lesssim 170 \Rsun$,
i.e., blue supergiants and relatively compact red supergiant
progenitors.  Such flashes are characteristically harder and dimmer
than those from more extended RSG progenitors; nevertheless, they are
visible to distances of several hundred Mpc and should be observed in
the hundreds per year (Table \ref{t:rates}).  

For these stars, the flash colour temperature $\Tse$ and
absorbing column $\NH$ can be estimated for extinctions $\NH\lesssim
2\times 10^{21}\ {\rm cm^{-2}}$ by placing them on duration-colour
(Figs. \ref{f:tvsC32} and \ref{f:tvsC21}) and colour-colour
(Fig. \ref{f:C-C}) diagrams.  $\Tse$ provides a constraint on the
explosion energy and ejected mass through the equations derived by
MM~99 (Table \ref{t:MM99eqs}); however, these quantities cannot be
derived independently without additional information.  In the case
that optical followup provides a distance, this degeneracy can be
broken (Figure \ref{f:L-C}). 

Events that outlast {\em LOBSTER}'s dwell time are overrepresented because
they can be detected even if they do not begin while inside the field
of view (\S \ref{lobster}).  However they are also poorly
characterized, because only a lower limit is available for their
duration.  With optical follow-up they can still be placed on Figure
\ref{f:L-C}, allowing an estimate of the obscuring column and the
colour and luminosity of the breakout flash.

\subsection{Implications}\label{implications}

From the above analysis we draw several conclusions.  The \lobster
space observatory will provide a census of Type II supernova events
that is complete within $\sim 250$ Mpc (Table \ref{t:rates} and Figure
\ref{f:DmaxvsNh}) and up to columns $\NH \gtrsim 10^{21} {\rm
cm^{-2}}$.  The number of events expected in three years of \lobster 
observations is quite uncertain, but for a conservative estimate of 
host-galaxy extinction (Table \ref{t:ratesExt}) it ranges from $\sim 50$ 
to $\sim 600$ depending on the typical radius of the progenitor stars.

A population of dim Type II supernovae from blue progenitors is
suggested by SN 1987A and its analogues \citep{1988slmc.proc..112S,
1988slmc.proc..106F, 1988Natur.333..305Y, 1996ApJ...464..404S},
although its significance is controversial \citep{1989ApJ...347L..29V}.  
SN 1987A analogues should appear in the data as flashes whose duration 
is shorter than the dwell time of the instrument.  These events, whose 
properties are difficult to infer from optical observations due to the 
contribution of $^{56}$Co (\S \ref{RelnToOpt}), will provide strong 
constraints on stellar evolution at the point of core collapse.

\subsection{Caveats}\label{caveats}
The current work relies on the extrapolation of breakout flash
properties for a variety of stellar progenitors from only a pair of
numerical calculations, by means of the analytical scaling relations
given by MM~99; see \S \ref{method}.  It is possible that elements of
the radiation dynamics cause the overall shape of the breakout
spectrum to change across this range, which would introduce a
systematic error into our predictions of X-ray observables.  This can
only be tested with a more systematic survey of breakout properties,
preferably using multigroup radiation hydrodynamics simulations. 

One limitation of the MM~99 scaling relations is that they cover only
two possible forms of the outer density profile ($\rho\propto {\rm
depth}^n$): $n=3/2$, representing convective envelopes, and $n=3$,
representing radiative envelopes with constant opacity.  Other
possibilities should be considered.  For instance, convection is
inefficient near the surfaces of red supergiants, and this effects
shock propagation in the outermost layers (as MM~99 noted).  A
superadiabatic layer has a value of $n$ that is lower than the
adiabatic value.  On the other hand, the MM~99 analysis and our own
investigations show that the detailed structure of the outer envelope
plays a rather minor role in the breakout flash intensity, colour, and
duration (see also the Appendix).

\subsection{Speculation: Asymmetric Explosions}\label{speculations}

As noted above, the short breakout flashes that fit within the
\lobster dwell time have durations set by the star's light travel
time.  This makes it possible for asymmetries in those explosions to
affect the time dependence of their breakout flashes.  A spherically
symmetric explosion will exhibit a progression in brightness and
colour that represents the growth of the emitting area and the change
of its limb darkening as the observable portion of the breakout moves
from the front to the side of the star \citep[e.g.,][]{ens92}.  If the
explosion is sufficiently asymmetric, then this pattern will be
disturbed by the motion of the shock front across the face of the
star.  Unfortunately, the most compelling evidence would derive from
an observation of time-dependent linear polarization of the emerging
X-rays; a difficult quantity to observe. 

Asymmetries can derive from several sources: asymmetries in the
envelope distribution (due to rotation and convective eddies); the
growth of (weakly) unstable perturbations in accelerating shocks; and
the asymmetry of the central engine driving the explosion, e.g., in
the case of a jet-driven explosion.  (Note that gamma-ray burst
jets are not likely to escape supergiant stars
[\citealt{2003MNRAS.345..575M}]; however they may
imprint an asymmetry on the explosions.)

Concentrating on the first of these, how much asymmetry would be
required to perturb the shock travel time by an amount comparable to
the light travel time?  The shock speed is roughly $1 (10
\Msun/\Mej)\%$ of $c$ on average, so a relative difference in travel time of
the same order is required; this would arise from a comparable
asymmetry in the progenitor.  This degree of asymmetry may exist in
blue supergiants if, for instance, they have undergone tidal
interactions with companion stars in previous red supergiant phases or
are close to their Roche radii at the time of explosion. 

\section*{Acknowledgements}

\noindent 
We thank the referee, Roger Chevalier, for pointing out that early SN
luminosity provides a constraint on stellar radius.  We are grateful to 
Nigel Bannister and the \lobster science team for their correspondence 
and for specifications of the instrument, and to Neil Brandt, David 
Ballantyne, and John Monnier for helpful suggestions.  AJC was supported 
in part by an NSERC undergraduate fellowship.  CDM is supported by NSERC 
and the Canadian Research Chairs Program.

\bibliographystyle{apj}

\appendix 

\section{Stellar outer density coefficients}\label{params}

The parameter $\rhorho$, which describes the density structure of the
outermost regions of a progenitor star (\S \ref{physics}), appears in
the MM~99 equations for the properties of the breakout flash.  Though
the near-surface structure plays little role in determining the
temperature and energy of a breakout flash, it is significant in
dictating the duration $\tse$ of a diffusion-dominated flash.

The following sections we estimate $\rhorho$ for the radiative
envelopes of blue supergiants and for the convective envelopes of red
supergiants.  In each case we motivate replacing $\rhorho$ with a
particular numerical value; only in exceptionally well-observed
breakout flashes could the dependence of $\rhorho$ on stellar
parameters be used to reconstruct them. 

\subsection{Outer Density Coefficients for Blue Supergiants}
\label{BSG}

It is possible to re-write the MM~99 scaling equations for BSGs (Table 
\ref{t:MM99eqs}) such that $\rhorho$ is eliminated in favour of the
mass, luminosity, and compositional parameters of the progenitor star.  
For progenitors with a radiative envelope (i.e., BSGs), $\rho_1$ may be 
expressed as

\begin{equation}\label{eq:radrho1}
\frac{\rho_1}{\rho_{\star}} = \frac{a (\mu m_H)^4}{192 k_B^4} 
\frac{G^3 M_\star^3}{\Mej} \frac{\beta^4}{1-\beta}, 
\end{equation}
where $1 - \beta \equiv L_\star / L_{\rm Edd}$ is the ratio of the stellar 
luminosity to the Eddington limit.  For Thompson opacity,

\begin{equation} 
1 - \beta = 0.259\left(\frac{L_{\star}}{10^5 L_{\odot}}\right)
\left(\frac{M_{\star}}{10 M_{\odot}}\right)^{-1}
\left(\frac{1+X_H}{1.7}\right), 
\end{equation} 
which leads to

\begin{eqnarray}\label{eq:noM-L}
\frac{\rho_1}{\rho_\star} &=& 0.571 \left(\frac{\mu}{0.62}\right)^4 
\left(\frac{M_{\star}}{10 \Msun}\right)^4 \nonumber\\ &\times& 
\left(1 - \frac{L_\star}{L_{\rm Edd}}\right)^4 
\left(\frac{\Mej}{10 \Msun}\right)^{-1} \nonumber\\ &\times& 
\left(\frac{L_{\star}}{10^4 L_{\sun}}\right)^{-1} 
\left(\frac{1 + X_H}{1.7}\right)^{-1}.
\end{eqnarray}

\begin{figure}
\centerline{\epsfig{figure=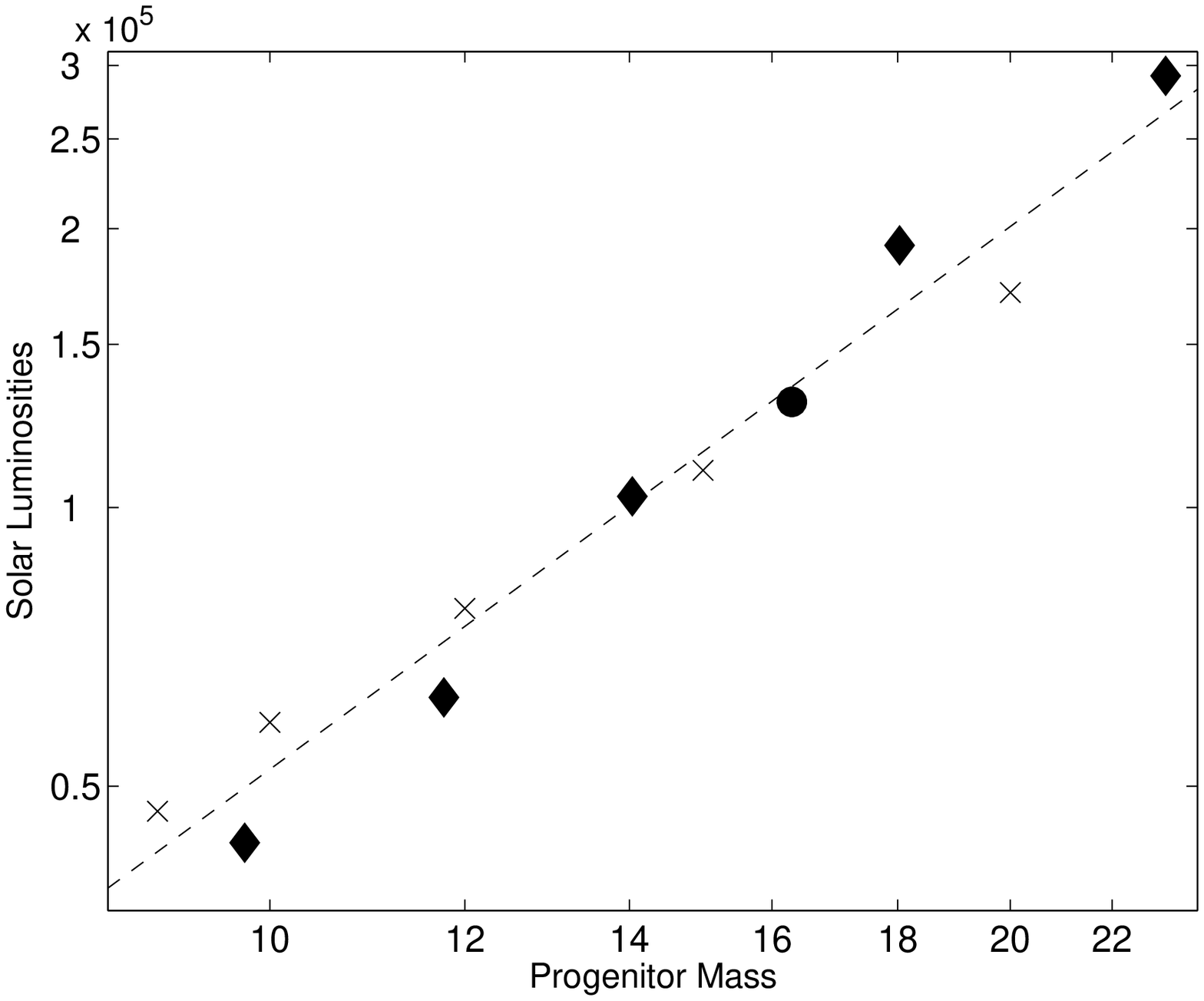,width=3.5in}} 
\caption[] {\footnotesize Luminosity-mass relation for presupernova stars.  
Plotted are the terminal masses and luminosities for the Padova tracks 
with solar-scaled metallicity $Z=0.008$ (Salachnich et al. 2000, A \& A 361, 
1023, crosses), for Geneva tracks of the same metallicity (Scherer et al. 1992 
A \& AS Supp. 98, 523), and for a progenitor model for SN 1987A (Hashimoto, 
Shigeyama, \& Nomoto 1989, circle).  Also plotted is a simple power law fit 
(Eq. \ref{LofM}). 
\label{f:M-L} }
\end{figure}

Using theoretical stellar model data from various authors, we have 
established a semi-empirical mass-luminosity (M-L) relation for 
presupernova stars.  This M-L relation, show in Figure \ref{f:M-L}, is 

\begin{equation}\label{LofM}
\frac{L_{\star}}{L_{\sun}} \approx 590 \left(\frac{M_{\star}}{\Msun}\right)^{1.95} 
\end{equation}
and it is valid for the mass range $10 < M_{\star} / \Msun < 20$.  
Substituting the M-L relation into equation (\ref{eq:noM-L}) results in:

\begin{eqnarray}\label{eq:rhorhoM-L}
\frac{\rho_1}{\rho_{\star}} &=& 0.107 \left(\frac{\mu}{0.62}\right)^4 
\left(\frac{M_{\star}}{10 \Msun}\right)^{2.05} \nonumber\\ &\times& 
\left(1 - \frac{L_{\star}}{L_{\rm Edd}}\right)^4 
\left(\frac{\Mej}{10 \Msun}\right)^{-1} \nonumber\\ &\times& 
\left(\frac{1 + X_H}{1.7}\right)^{-1}.
\end{eqnarray}

From equation (\ref{eq:rhorhoM-L}) it follows that $\rhorho$ varies
almost linearly from $\sim 0.13$ to $\sim 0.26$ for $10 < M_\star 
/ \Msun < 20$, assuming a $1.5 \Msun$ remnant and $L_{\star} << L_{\rm Edd}$.  
Thus, we will adopt the average, $\sim 0.2$, as a fiducial value for all BSG 
progenitors.

\subsection{Outer Density Coefficient for Red Supergiants}
\label{RSG}

Combining equations (9), (14), and (48) from MM~99 results in:
\begin{equation}
\frac{\rho_1}{\rho_\star} = \left(12.62 - 12.17q\right)
\left(0.37 - 0.18q + 0.096q^2\right)^{5/2},
\end{equation}
where $q \equiv 1 - M_{\rm env} / M_\star$ and $M_{\rm env}$ is the
mass of the outer stellar envelope.  In this case, $\rhorho$ 
exhibits an approximately $1/q$ dependence for the most plausible
range of $q$ values, $0.3 < q < 0.5$ (see Figure 5 of MM~99).  For
these values of $q$, $\rhorho$ varies from $\sim 0.54$ to $\sim 0.33$;
we adopt $0.5$ as the standard for RSGs.

\begin{table}\label{t:noM-L}
\centering
\caption{Alternate scaling equations for progenitors with 
radiative envelopes, without using the M-L relation. {These should 
be read as exponents in formulae, e.g., $\tse = 10^{1.65}[f\kappa/(0.34\ 
{\rm cm^2 g^{-1}})]^{-0.14}$... cgs units.} }
\begin{tabular}{@{}lrrr@{}}
\hline
Parameter & $\Tse$ & $\Ese$ & $\tse$ \\ 
\hline
$10$ & $6.11$ & $47.37$ & $1.65$ \\
$f\kappa/0.34$ cm$^2$ g$^{-1}$ & $-0.14$ & $-0.84$ &$-0.45$ \\
$\Ein/10^{51}$ erg & $0.18$ & $0.58$ & $-0.72$ \\
$\Mej/10 \Msun$ & $-0.11$ &  $-0.36$ & $0.46$ \\
$\Rstar/50 \Rsun$ & $-0.48$ & $1.68$ &$1.90$ \\
$M_{\star}/10 \Msun$& $0.18$ &  $-0.22$ &$-0.73$ \\
$1 - L_{\star}/L_{\rm Edd}$ & $0.18$ & $-0.22$ & $-0.73$ \\
$\mu/0.62$ & $0.18$ & $-0.22$ & $-0.73$ \\
$(1 + X_H)/1.7$ & $-0.046$ & $0.054$ & $0.18$ \\
$L_{\star}/10^4 L_{\sun}$ & $-0.046$ & $0.054$ & $0.18$ \\
\hline
\end{tabular}
\end{table}

\begin{table}
\centering
\caption{Alternate scaling equations for progenitors with radiative
envelopes, using the mass-luminosity relation
(Eq. [\ref{LofM}]). These should be read as exponents in formulae,
e,g., $\tse = 10^{1.78}[f\kappa/(0.34{\rm cm^2
g^{-1}})]^{-0.45}$... cgs units. \label{t:M-L}}

\begin{tabular}{@{}lrrr@{}}
\hline
Parameter & $\Tse$ & $\Ese$ & $\tse$ \\
\hline 
$10$ & $6.08$ & $47.41$ & $1.78$ \\
$f\kappa/0.34$ cm$^2$ g$^{-1}$ & $-0.14$ & $-0.84$ &$-0.45$ \\
$\Ein/10^{51}$ erg & $0.18$ & $0.58$ & $-0.72$ \\
$\Mej/10 \Msun$ & $-0.11$ &  $-0.36$ & $0.46$ \\
$\Rstar/50 \Rsun$ & $-0.48$ & $1.68$ &$1.90$ \\
$M_{\star}/10 \Msun$& $0.10$ &  $-0.12$ &$-0.40$ \\
$1 - L_{\star}/L_{\rm Edd}$ & $0.18$ & $-0.22$ & $-0.73$ \\
$\mu/0.62$ & $0.18$ & $-0.22$ & $-0.73$ \\
$(1 + X_H)/1.7$ & $-0.046$ & $0.054$ & $0.18$ \\
\hline
\end{tabular}
\end{table}

\end{document}